# Simple, Flexible, and Interoperable SCADA System Based on Agent Technology

**Hosny Abbas[1]\*, Samir Shaheen[2], Mohammed Amin[1]**

[1]Department of Electrical Engineering, Assiut University, Assiut, Egypt
[2]Department of Computer Engineering, Cairo University, Giza, Egypt
Email: \*hosnyabbas@aun.edu.eg, mhamin@aun.edu.eg, sshaheen@eng.cu.edu.eg

## Abstract

SCADA (Supervisory Control and Data Acquisition) is concerned with gathering process information from industrial control processes found in utilities such as power grids, water networks, transportation, manufacturing, etc., to provide the human operators with the required real-time access to industrial processes to be monitored and controlled either locally (on-site)or remotely (*i.e.*, through Internet). Conventional solutions such as custom SCADA packages, custom communication protocols, and centralized architectures are no longer appropriate for engineering this type of systems because of their highly distribution and their uncertain continuously changing working environments. Multi-agent systems (MAS) appeared as a new architectural style for engineering complex and highly dynamic applications such as SCADA systems. In this paper, we propose an approach for simply developing flexible and interoperable SCADA systems based on the integration of MAS and OPC process protocol. The proposed SCADA system has the following advantages: 1) simple (easier to be implemented); 2) flexible (able to adapt to its environment dynamic changes); and 3) interoperable (relative to the underlying control systems, which belongs to diverse of vendors). The applicability of the proposed approach is demonstrated by a real case study example carried out in a paper mill.

## Keywords

## 1. Introduction

A SCADA system is responsible for gathering information and real-time data from variety of plants and provid-ing this data to operators located at anywhere at any time. Furthermore, SCADA systems can be considered as critical information systems; their criticality comes from the fact that they are currently vital components of most nations' critical infrastructures. They control pipelines, water and transportation systems, utilities, refineries, chemical plants, and a wide variety of manufacturing operations. Failure of controlled systems can lead to direct loss of life due to equipment failure or indirect losses due to failure of critical infrastructure controlled by SCADA. SCADA systems have evolved in parallel with the growth and sophistication of modern computing technology. That means that SCADA is technology-dependent and it is not a standalone science but it is the result of integrating variety of applied sciences such as communication, computers, software engineering, networking, security, etc. Therefore and according to the evolution of technology, SCADA developers architect SCADA by selecting the appropriate technologies and mechanisms for handling SCADA challenges such as complexity resulted from the continuous increasing of the size of SCADA systems. SCADA systems should be scalable because their components and the amount of exchanged data increase with rapid rate. Further, a SCADA system should be flexible and able to adapt to internal or external changes. SCADA is no longer restricted to concern real-time monitoring and control inside a factory, but it currently has new concerns. It has been moved from the local sense to the global one. For example, in advanced countries, a global SCADA system is used for supervising and real-time monitoring of the power grid (thousands of electric power generation and distribution stations).

\*Corresponding author.

Designing, monitoring and controlling of modern industrial systems is getting more challenging as a consequence of the steady growth of their size, complexity, level of uncertainty, unpredictable behavior, and interactions. Unfortunately, the conventional SCADA systems are not capable of providing information management and high-level intelligent approaches. That is because achieving those functionalities requires comprehensive information management support and coordination among system devices, and the control of many different types of task, such as data transportation, data display, data retrieval, information interpretation, control signals and commands, documentation sorting and database searching…,etc. These tasks operate at different timescales and are widely distributed over the global system and its subsystems. Without sophisticated software architectures and hardware structures, it is impossible to handle these tasks efficiently, safely and reliably, with the possibility of online reconfiguration and flexibly embedding applications [1] [2].

It becomes obvious that computing systems, especially those related to modern industrial applications such as SCADA systems, are becoming increasingly interconnected and more difficult to maintain. Due to the increase in the size, complexity and the number of components, it is no longer practical to anticipate and model all possible interactions and conditions that the system may experience at design time. Similarly, the systems are becoming too large and too complex for system managers to maintain them at run-time [3].

The agent-based approach seems to be the promising solution. The rapid development of the field of agent-based systems offers a new and exciting paradigm for the development of sophisticated programs in dynamic and open environments [4]. The agent-based approach is considered as a new software engineering architectural style for the development of complex, decentralized and open software applications. What distinguishes the agent-based approach from other traditional approaches is its unique ability to handle simultaneously many challenges of the current software applications specially those applications which are highly distributed and their working environments are highly dynamic and uncertain. MAS provide a suitable paradigm for decentralized systems in which autonomous individuals engage in flexible high-level interactions.

By an agent-based system, we mean the one in which the key abstraction used is that of an agent. We therefore expect an agent-based system to be both designed and implemented in terms of agents. An agent-based system may contain any non-zero number of agents. The multi-agent case where a system is designed and implemented as several interacting agents, is both more general and significantly more complex than the single-agent case. Originally, MAS emerged as a scientific area, from the previous research efforts in Distributed Artificial Intelligence (DAI) started in the early eighties. MAS are now seen as a major trend in R & D, mainly related to artificial intelligence and distributed computing techniques. This research has attracted attention in many application domains where difficult and inherently distributed problems have to be tackled [5]. A MAS is defined as a set of interacting autonomous agents in a common environment in order to solve a common, coherent task. These agents try to achieve individual objectives which are sometimes conflicting. There are many definitions of the meaning of agent we found in their literature but the widely accepted definition of an agent is that which has been stated by Wooldridge and Jennings [6] who defined an agent as:

"*... a hardware or* (*more usually*) *software-based computer system that enjoys the following properties*: *–autonomy: agents operate without the direct intervention of humans or others, and have some kind of control over their actions and internal state*; *–social ability: agents interact with other agents* (*and possibly humans*) *via some kind of agent-communication language*; *–reactivity: agents perceive their environment,* (*which may be the physical world, a user via a graphical user interface, a collection of other agents, the Internet, or perhaps all of these combined*), *and respond in a timely fashion to changes that occur in it*; *–pro-activeness: agents do not simply act in response to their environment, they are able to exhibit goal-directed behavior by taking the initiative*".

A MAS is autonomous, means that there is no external entity which controls this system. This property is enforced because agents inside the system are autonomous. Inside a MAS, data (knowledge) are distributed inside all its agents. Moreover, the control is decentralized (there is no supervisor). MAS allow the design and implementation of software systems by using the same ideas and concepts that are the very founding of human societies and habits. These systems often rely on the delegation of goals and tasks among autonomous software agents, which can interact and collaborate with others to achieve common goals [7]. It provides an approach to solve a software problem by decomposing the system into a number of autonomous entities embedded in an environment in order to achieve the functional and quality requirements of the system [8]. The agent-based computing has been hailed as "the next significant breakthrough in software development" [9], and "the new revolution in software" [10].

Currently, agents are the focus of intense interest on the part of many sub-fields of computer science and artificial intelligence. Agents are being used in an increasingly wide variety of applications, ranging from comparatively small systems such as email filters to large, open, complex, mission critical systems such as air traffic control. At first sight, it may appear that such extremely different types of system can have little in common. And yet this is not the case: in both, the key abstraction used is that of an agent [11]. MAS are claimed to be especially suited to the development of software systems that are decentralized, can deal flexibly with dynamic conditions, and are open to system components that come and go. That is why they are used in domains such as manufacturing control, automated vehicles, and e-commerce markets.

There are many reviews and surveys tackled with the applications of agent and multi-agent systems in variety of application domains. For instance, in [11] the authors described the suitability of intelligent and autonomous agents to model and develop applications for certain types of software system which are inherently more difficult to correctly design and implement than others. And they subdivide these systems into three classes: open systems, complex systems, and ubiquitous computing systems. Moreover, they classified these systems or applications according to the application domain such as industrial applications (*i.e.* manufacturing, process control…, etc.), commercial applications (*i.e.* information management, e-commerce…, etc.), medical applications (*i.e.* patient monitoring, health care…, etc.), and entertainment (*i.e.* games, interactive cinema…, etc.). The authors tackled many types of agent-based applications and described generally their functionality. They tackled the design and development methods and challenges of MAS from a general viewpoint independent from a specific application domain.

In [12] the author presented a summary of the state-of-the-art of the Distributed Artificial Intelligence (DAI) applied to Intelligent Manufacturing and presented main applications along with different technologies applied in these areas. Also, he presented areas of agent technology applications and the agent development tools. Moreover, he described briefly and generally the agents and multi-agent architecture types and design approaches.

In [13] the authors reviewed multi-agent systems power engineering applications and stated that the flexibility offered by an open architecture of agents with good social ability easily led to the design of a fault-tolerant system. Also, they reviewed the application of MAS as a modeling and simulation approach and its application in grid computing and web services composition, etc. Moreover they presented a bibliographical analysis of agent research aiming to provide an indication of the active areas of agent research with respect to power systems and related applications. They concentrated on the functionalities and the different power systems aspects which can be designed and developed by MAS.

Shehory [14] stated that one aspect of multi-agent systems that had been only partially studied was their role in software engineering and especially their merit as a software architecture style. In his report he provided analysis guidelines which supported designers in their assessment of the suitability of MAS as a solution to computational problem they addressed. Moreover he discussed the architectural properties that should be consi-dered when analyzing such systems and he supported his work with case studies of several MAS.

Still there is a debate about the identity of multi-agent systems: is it a radically new way for systems engineering? Is it a new software engineering modeling style? Weyns *et al.* [8] stated that the trend in agent-oriented software engineering was to consider multi-agent systems as a radically new way of engineering software, and this position isolated agent-oriented software engineering from mainstream software engineering and could be one important reason why MAS were not widely adopted in industry yet. What we understand from this is that the agent-based modeling should be considered similar to the object-oriented modeling both of them are software engineering modeling styles. In other words, MAS are considered now as a novel general-purpose paradigm for software development.

Moreover, Agents and multi-agent systems constitute one of the most prominent and attractive technologies in computer science at the beginning of this new century. Agents and multi-agent systems technologies, methods, and theories are currently contributing to many diverse domains. Jennings and Wooldridge [11] stated that intelligent agents were a new paradigm for developing software applications. Agent technology is seen as a fundamentally important new tool for building a wide array of systems (*i.e.* open systems, complex systems, and ubiquitous computing systems). A number of software tools exist that allow a user to implement software systems as agents, and as societies of cooperating agents, by tools we mean agent platforms. An agent platform provides a basis for the implementation of MAS, and the means to manage agent execution and message passing. For the sake of interoperability, it is intended that the agent platform architecture should be implemented by using the Foundation of Intelligent and Physical Agents [15] specifications of agent platforms abstract architecture. FIPA is an IEEE Computer Society standards organization that promotes agent-based technology and the interoperability of its standards with other technologies. FIPA, the standards organization for agents and MAS was officially accepted by the IEEE as its eleventh standards committee on June 8, 2005. The specifications define an abstract agent platform, a number of services that must or may be provided by such a platform and a standard communications language.

From the above discussion it is clear that the agent technology can offer a feasible approach for handling the challenges and requirements of modern SCADA systems [16]. This paper proposes a simple but flexible and interoperable SCADA system based on the agent-based approach. The proposed system-to-be can be used for providing not only a local access but also a remote access through the Internet to local control processes. The remaining of this paper is organized as follows: Section 2 provides a short background of the concerned problem; Section 3 presents the proposed practical approach; Section 4 presents the real case study application of the proposed practical approach; Section 5 concerns the deployment and testing of the proposed SCADA system; and finally, Section 6 concludes the paper and highlights the future intentions.

## 2. Background

Traditionally, SCADA developers adopted the web-based approach to realize a remote real-time monitoring of production control processes, a survey of some web-based SCADA systems can be found in [17]. But the problem of the web-based applications which use the web browser as a remote client and a web server as the onsite server through which the remote operators have access to control processes, is an example of centralized control architecture which has disadvantages such as:

1) Single point of failure problem;
2) These systems are not flexible because they cannot adapt to changes such as increasing the number of system components;
3) These systems are not scalable because increasing system components can result in web server overload and crash.

Figure 1 demonstrates why the web-based SCADA systems are not flexible, not scalable, and are not able to adapt to working environments dynamic changes. As shown in Figure 1(a) a remote operator uses the web browser to connect to an onsite web server to access a control process. In case the system changes by increasing the number control processes and the number of remote operators as shown in Figure 1(b) the web server which has limited resources such as the computation power, memory and the network bandwidth, is vulnerable to be overloaded and crash.

Many of SCADA developers considered the web-based SCADA systems as a solution to some of modern SCADA challenges and shortcomings [17] [18]. Although they succeeded to develop web-based SCADA applications that had a number of important quality attributes such as efficiency, interoperability, and real-time monitoring, they failed to obtain other important attributes such as scalability and adaptivity. The reason is that the Hypertext Transfer Protocol (HTTP) uses a client-server architecture with which it is difficult to achieve higher scalability degree because of the dominant adopted centralized control regime. The author of [19] emphasized that the future infrastructures will contain a huge amount of data is generated by real world devices and needs to be integrated and processed within a specific context and communicated on demand and on time. As a result, traditional approaches aiming at the efficient data inclusion in enterprise services need to be changed. The main challenges facing current and future SCADA systems are generally related to quality attributes. The extent to which the system possesses a desired combination of quality attributes such as scalability, usability, performance, reliability, and security indicates the success of the design and the overall quality of the software system. New design and development approaches should be adopted for building future complex and critical SCADA systems.

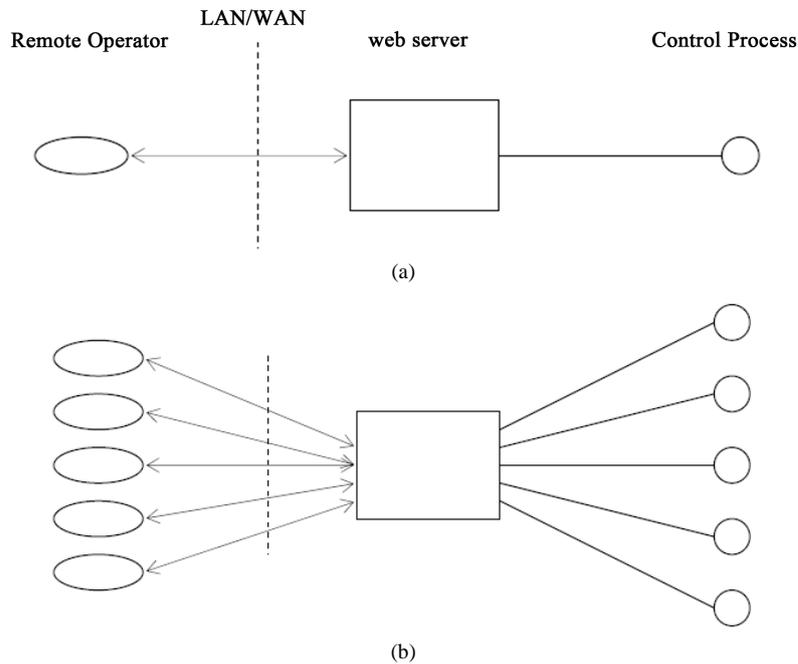

**Figure 1.** Traditional centralized web-based SCADA architecture.

The agent-based approach considers a SCADA system as a multi-agent system. A multi-agent system provides a powerful computational technology, for which dynamic aspects are based on interactions between autonomous agents, rather than centralized control. To do their required functions, agents have the ability to communicate with each other, with their environment, and with human operators. Agents are particularly adapted to complex systems modeling where environments are unpredictable [20]. **Figure 2** provides an abstract architecture of the agent-based real-time monitoring. As shown, the system is flexible and can easily adapt to changes such as increasing the number of control processes which can easily assigned to new agents, and also increasing the number of remote operators by assigning a remote agent to each operator. The agent-based approach adopts a decentralized control and behavior, which had proven to be the appropriate approach for handling system complexity and unpredictable work environments.

The agent-based approach can be considered as the promising solution for the design and development of future SCADA because it has the ability to handle future SCADA challenges such as complexity, scalability, and flexibility, etc. Although these quality attributes are not unique to MAS, but combining them in a single system is unique to MAS. This combination results in the suitability of MAS for solving problems where information, location and control are highly distributed, heterogeneous, autonomous components comprise the system, the environment is open and dynamically changing, and uncertainty is present [14] [21].

The agent-based approach reinforces decentralization which is the effective way for obtaining higher degree of scalability. Today, agents are being applied in a wide range of industrial applications, for example, process control, manufacturing, air traffic control, etc. These applications have been relatively successful, suggesting that the multi-agent approach is a promising method for the implementation of industrial automation systems and that encourages us to adopt MAS for the design and development of SCADA systems. Metzger *et al*. [22] surveyed the applications of the agent technology in the industrial process control and concluded that the agent technology is particularly popular in the manufacturing domain, while the applications in other domains of industrial control are scarce. They related their conclusions to the lack of the technology support on the part of control instrumentation vendors. In manufacturing automation, the process consists of discrete and countable components and actions. The natural approach is to assign the software agents to each of the components and each of the actions performed. On the other hand, the process automation deals with the continuous physical phenomena, such as chemical reactions. When a process automation system is designed, the phenomena are represented as mathematical models, for which control algorithms are chosen in order to keep the process parameters within a desired range. Therefore, in a single continuous control loop, there is not much place for any additional computational techniques, including the agent technology.

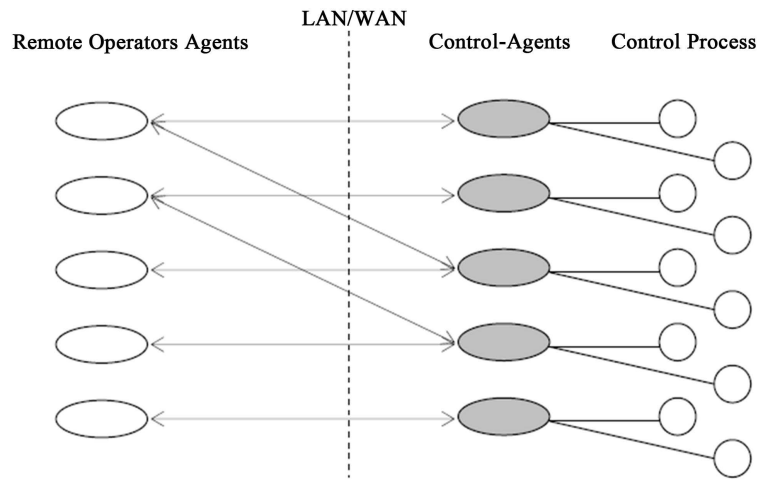

**Figure 2.** The decentralized agent-based SCADA.

## 3. The Proposed Approach

The proposed approach is based on the integration of MAS and OPC process protocol [23], realizing this integration enables us to achieve two goals. First, it will be possible to transfer the process data from the process domain to the information domain (MAS), where executive management can use this data for decision making. Second, it will be possible to take the benefit of control devices interoperability provided by the OPC process protocol. Interoperability is assured through the creation and maintenance of non-proprietary open standards specifications. OPC initially meant *Ole for Process Control*, but after it becomes familiar for achieving control systems interoperability it was redirected to mean *Open Process Control*. OPC is open connectivity in industrial automation and the enterprise systems that support the industry. The first OPC standard specification resulted from the collaboration of a number of leading worldwide automation suppliers working in cooperation with Microsoft. Originally based on Microsoft's OLE COM/DCOM technologies, the specification defined a standard set of objects, interfaces and methods for use in process control and manufacturing applications to achieve interoperability. There are now hundreds of OPC Data Access (OPC DA) servers and clients.

**Figure 3** shows the basic architecture of the OPC protocol usage. The OPC server is connected to the PLC (programmable Logic Controller) which is responsible of directly controlling a certain control process. By this way the OPC server can be considered as the driver of the PLC, in the same sense as the printer driver. The communication between the OPC server and the PLC is vendor-specific and depends on the custom technologies used by the vendors for manufacturing their control systems. On the other hand, the communication between the OPC server and software applications adopts the COM/DCOM technologies created by Microsoft. That is considered as a limitation because it means that the OPC protocol is bonded only to Microsoft Windows operating systems. In this research and as shown in figure a Siemens S7-400 PLC is used as a real control system but the same system and approach can be used with other control systems manufactured by other vendors such as ABB,

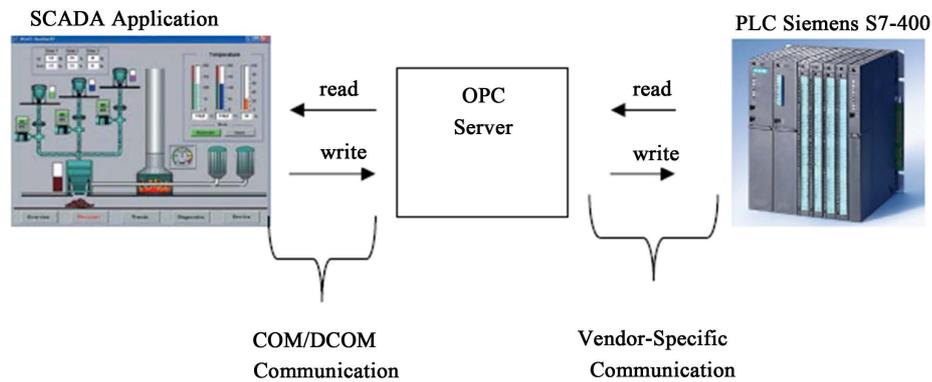

**SCADA Application**

read

OPC Server

read

write

write

**PLC Siemens S7-400**

COM/DCOM Communication

Vendor-Specific Communication

**Figure 3.** Basic OPC protocol architecture.

Honeywell, etc. That is possible because of the vendors interoperability provided by the standard OPC protocol. This feature represents the second goal (interoperability) of the aimed two goals mention in the beginning of this section.

Most MAS platforms are implemented in Java programming language [24], therefore to establish a connection between a Java agent and an OPC server, a Java-COM bridge or adapter is required as shown in **Figure 4**. For the purpose of this thesis the *JEasyOPC* Java-OPC adapter [25] [26] was selected to interface a Java MAS agent with OPC servers and of course the other bridges can be used in a similar way. *JEasyOPC* is a Java OPC client that is now greatly enhanced. It uses a JNI layer coded in Delphi. The current version supports both OPC DA 2.0 and OPC DA 3.0. *JEasyOPC* is free source and can be downloaded easily from the Internet. **Appendix A** provides some information about realizing the Java-OPC interactions through JEasyOPC.

**Figure 5** demonstrates the layered nature of the system-to-be. The system contains four layers (or worlds) with three interfaces in between. The layers are (from bottom to up): the physical process layer, control systems layer, OPC communication layer, and the agents' layer which represents the application world. The figure also demonstrates the appropriate interface between any two layers. For example, the interface between the physical process and its control system is the field interface, which is constructed using analog/digital signals transferred through wires, or it can be a digital field bus. Further, the interface between the control system and the OPC driver layer is vendor specific. Finally, the interface between the OPC layer and the agents' world layer a COM/ DCOM interface. In this paper, we are interested in the later (top) layer and interface (Shown in Dashed line in **Figure 5**).

To connect non-agent Java applications to OPC servers, it is required to download the *JEasyOPC* library, which includes a sample Eclipse [27] java project that contains basic sample examples programmed to connect to a default OPC server (Matrikon. OPC. Simulation). These examples can be modified as required, for instance the Makitron OPC server can be replaced with a Siemens one (*i.e.* OPC. Simatic Net). Furthermore, the examples don't provide a graphical user interface (GUI) and therefore the developer has to modify them to provide a suitable GUI. Before running Eclipse, a recent Java run-time environment (JRE) should be installed on the operating system. Using a Java development environment such as Eclipse frees the developer from caring about modifying related system variables such as CLASPATH and PATH as it does these issues automatically. From the other hand, to connect a Java agent (under Eclipse) to an OPC server, it is required first to install a MAS platform such as Jade (Java Agent Development Environment) [28]. Jade is a software framework fully implemented in Java language. It simplifies the implementation of multi-agent systems through a middleware that claims to comply with the FIPA specifications and through a set of tools that supports the debugging and deployment phase. The agent platform can be distributed across machines with different operating systems and the configuration can be controlled via a remote GUI. The configuration can be even changed at run-time by creating new agents and moving agents from one machine to another one as and when required. The only system requirement is the Java Run-Time version 5 or later. JADE is distributed in open source. To run Jade under Eclipse, the developer should add Jade libraries to Eclipse Java build path (*project → prosperities → Java Build path → Libraries → add external Jars*), then through the Windows file system find *Jade.jar* file in the Jade home as shown in **Figure 6**. Now Eclipse is ready for creating a new java class that extends *jade. core. Agent* class and start programming the required agent let us call it OPC-Agent.

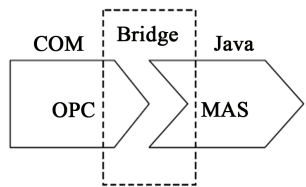

**Figure 4.** Integrating MAS with OPC protocol.

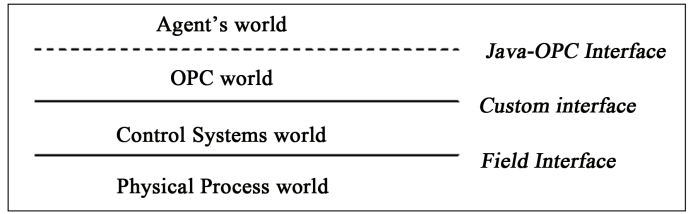

**Figure 5.** System layers and the interfaces in between.

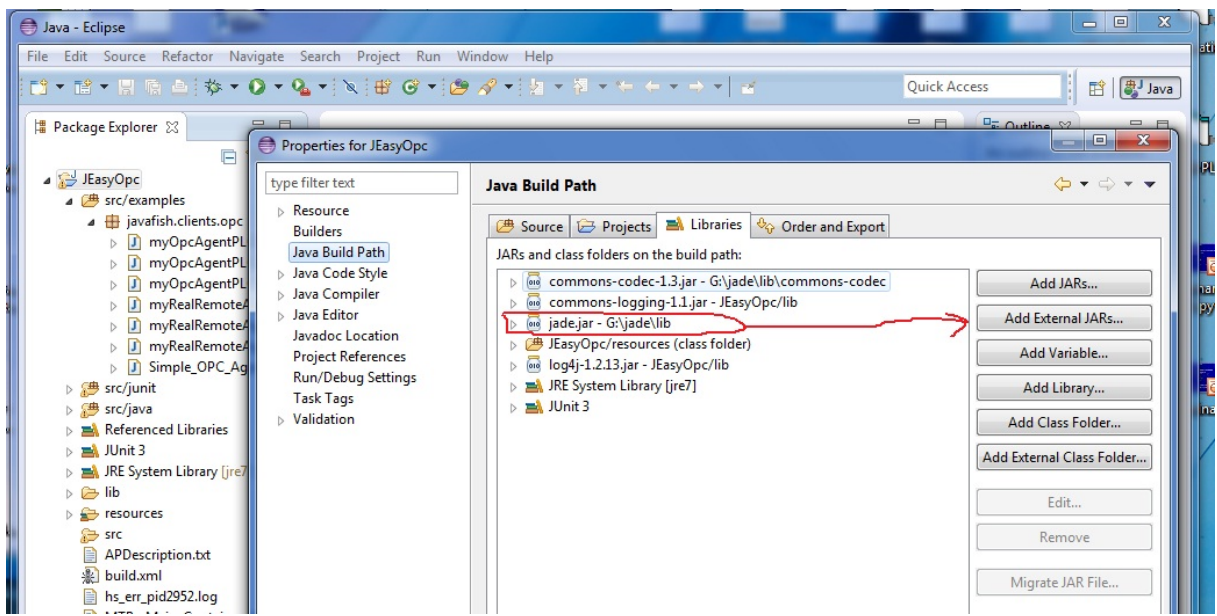

**Figure 6.** Running JADE from ECLIPSE.

## 4. Real Case Study Example

The case study is concerned with the real adoption of the agent-based approach for connecting locally and remotely to real control processes. The case study was executed in a paper mill, specifically in the finishing area within a paper mill. The paper mill finishing area comprises three separated stations as follows:

1) The Winder station, which takes a 6 meters width paper spool as its input and gives a smaller width paper rolls (*i.e.*, 6 × 1 meter) as output.
2) The wrapping station, which is used for paper rolls packaging and labeling.
3) The Salvage winder station for preparing sample rolls; it is a smaller Winder station.

Each control process is controlled by a Siemens S7-400 PLC; also the three stations are connecting a local LAN including the operator stations and the control systems. Further, each control process contains many process variables that should be monitored continuously, for instance, for the winder station the following variables required to be monitored in real-time: Machine Speed (m/min), Paper Tension (N/m), Drums Torques (N/m), and so on. **Figure 7** shows the winder station while it is running. The architecture of the experimental practical project is shown in **Figure 8**. As shown in the figure, all hosts are connected through a LAN with the possibility to be connected to the Internet to enable real-time data accessing from outside the paper mill. Furthermore, the normal remote Jade agents are considered as the process real-time data consumers while each one can be hosted on a different host in the LAN or in the Internet. In this case study project, it is assumed that:

1) Each PLC of a control process is assigned to an OPC-Agent, which in turn registers its services to the JADE DF (Directory Facilitator) agent, represents the yellow page service in the JADE platform. Therefore, in the case study example there are three OPC-Agents, one for each real PLC system.

2) The remote operator agents initially don't know which OPC-Agent is responsible of which PLC system. Accordingly, after start running they contact the JADE DF agent to search for the required OPC-Agents, as illustrated in **Figure 9**.

3) After identifying the required OPC-Agents, the operator agents send an ACL message with per-formative (Request) to the OPC-Agent asking her to continuously send back the real-time process data of its assigned PLC. This request is done only once then the OPC-Agent will continuously send the required real-time process data through ACL messages with per-formative (Inform), this mechanism is called subscription behavior.

4) Each operator agent can access many OPC-Agents simultaneously, in other words each remote agent can be used to supervise and monitor more than one control process.

5) The project JADE platform was named as SCADA, so an agent with local name "H1" has a full name "H1@SCADA".

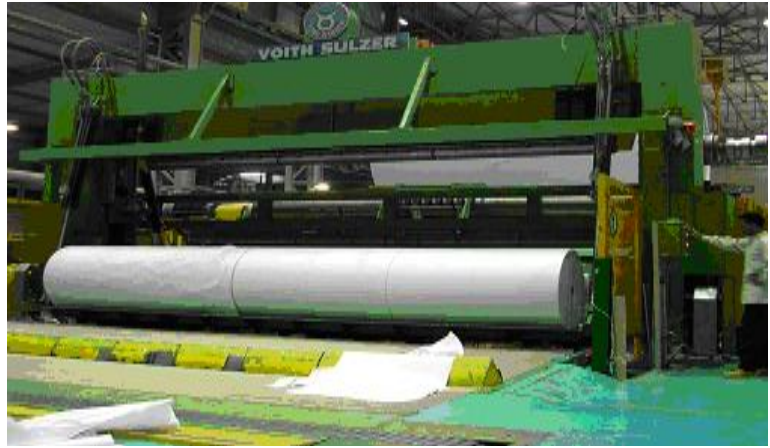

**Figure 7.** Qena paper winder station.

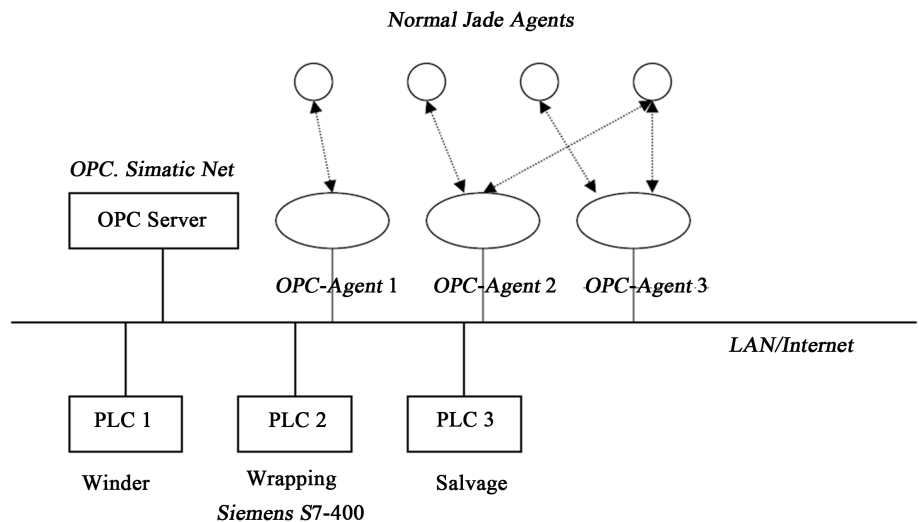

**Figure 8.** Architecture of the concrete example.

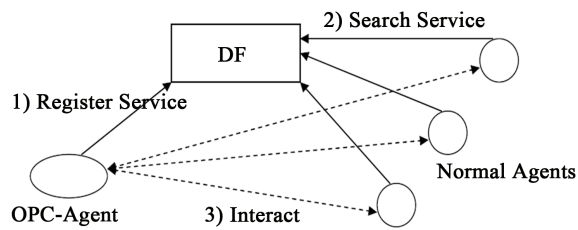

**Figure 9.** Exploiting JADE yellow page service.

6) The implemented agents were made simple with light GUI because it is just an illustrative experimental example. In real applications, the agents GUI may become complex and user friendly. **Figure 10** shows a simple remote operator agent continuously receives changed data from an OPC-agent situated on a host inside the mill. The remote agent designed to provide real-time monitoring, alarm service, and trend service, as shown in the figure. This feature represents our first goal (remote real-time monitoring) as we mentioned in the beginning of this section. This feature is very important for executive management stuff, which might be located far from the factory to be able to take the suitable quick decision in the suitable time.

It is possible to run an operator agent from any host in the LAN or even from the Internet. The point will be with the agents launching commands which defined in the Eclipse class arguments setting (*run as→run configuration → new configuration → arguments*), as shown in **Figure 11**. For example (considering *javafish. clients.opc* as the package name):

- For initially booting and running an OPC-Agent called H1 in the main container of the JADE platform called SCADA:
  - *-gui-name SCADA -local -agents H1: javafish.clients.opc.myOpcAgent*
- For running an operator agent called R1 on the same host as the OPC-Agent H1:
  - *-container-agents R1:javafish.clients.opc.myRealRemotAgent1*
- For running an operator agent called R2 from another host in the LAN Where the host with IP (192.168.100.31) is the one on which the Jade platform main container is situated:
  - *-container-host 192.168.100.31 R2:javafish.clients.opc.myRemoteAgent*

For debugging purposes, the JADE platform provides a built-in sniffer tool, which is used to visualize messages as a low-level UML sequence diagram, was used as shown in **Figure 12**, which demonstrates how JADE agents interacts together through exchanging asynchronous messages. The sniffer tool is very powerful for tracing the system-to-be behaviors in run-time.

Returning to the concerned case study, **Figure 13** presents the system sequence diagram. As shown in the figure the remote operator agent depends on the yellow page service provided by the JADE directory facilitator (DF) to find the control agent responsible of a required control process. The operator agent will wait for the DF until it provides it with the required control process. After getting a response from the DF the remote agent starts interacting with the control agent to get the process real-time data and presenting this data to the human operator in text or graphical format. As we have illustrated before the remote operator agent can be hosted locally onsite or remotely outside, *i.e.* using the Internet.

## 5. Deployment and Testing

The proposed approach solves the problem of accessing the OPC server (which is a COM/DCOM server) from the Internet because using DCOM communication is not recommended through the Internet because of its delay time and configuration security problems. The agent-based approach depends on the HTTP protocol, which is a firewall-friendly protocol. **Figure 14** shows the concrete project while it is running. The figure includes two running OPC-Agent in the left side:

- WinderOpcAgent 1;
- WrappingOpcAgent 1;
  And three remote normal agents in the right side:
- WinderRemoteAgent 1;
- WinderRemoteAgent 2;
- WrappingRemoteAgent 2;
  **Figure 15** demonstrates how the OPC-Agents continuously send real-time data to the remote operator agents as a result of subscription requests from the operator agents. The figure shows two OPC-Agents and three remote agents. Two remote operator agents subscribed to one of the two OPC-Agents and one subscribed to the other one. The number of OPC-Agents and remote agents can be increased easily by instantiating these agent types as required and anywhere in the LAN or the Internet. Note that it's not necessary for the OPC-Agents to have a GUI but the operator agents should have a user friendly GUI. In this experiment, a simple GUI is designed for both agent types.

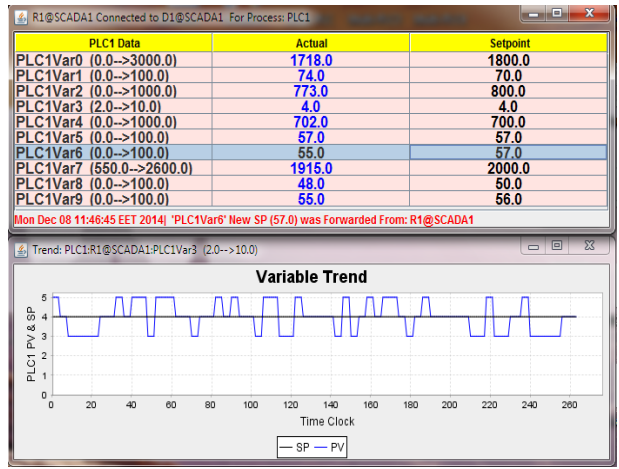

**Figure 10.** A remote operator agent connected to an onsite OPC-agent and provides real-time process monitoring, Alarm service, and Trend service.

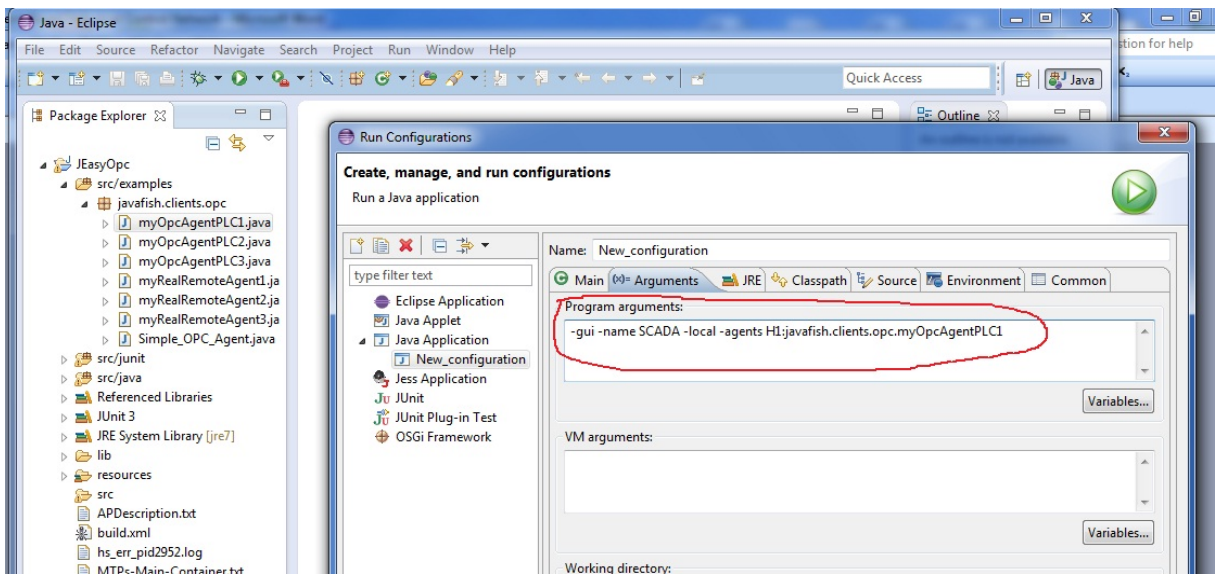

**Figure 11.** Configuring a JADE agent in Ellipse.

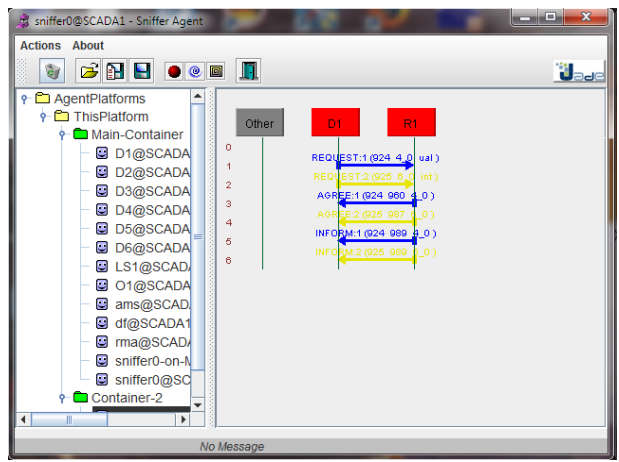

**Figure 12.** Jade sniffer tool shows message flow between two agents.

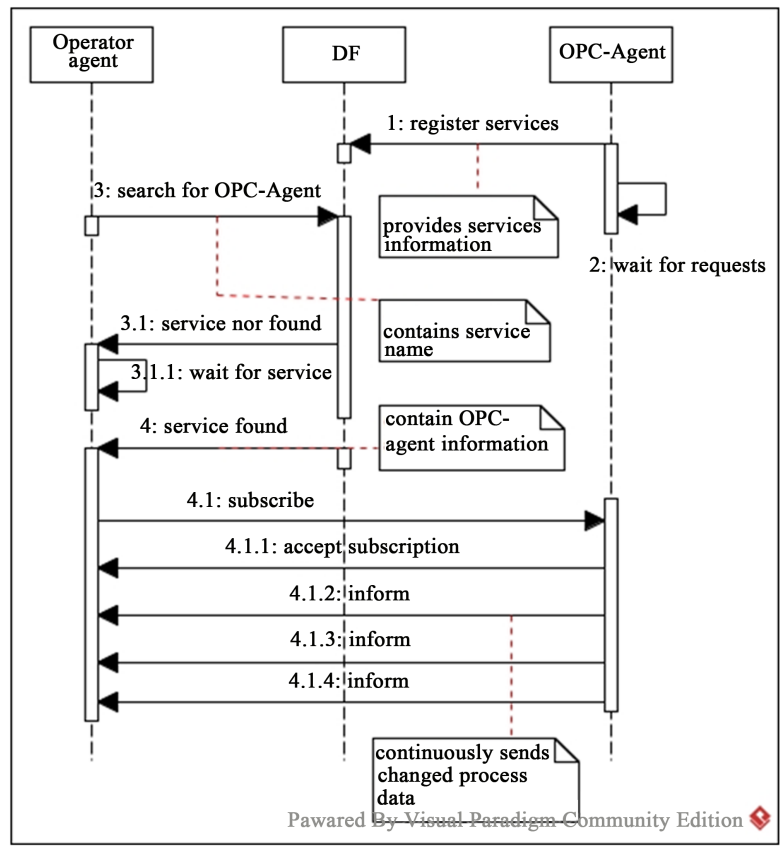

**Figure 13.** System sequence diagram demonstrated how a remote agent finds a control agent to interact with.

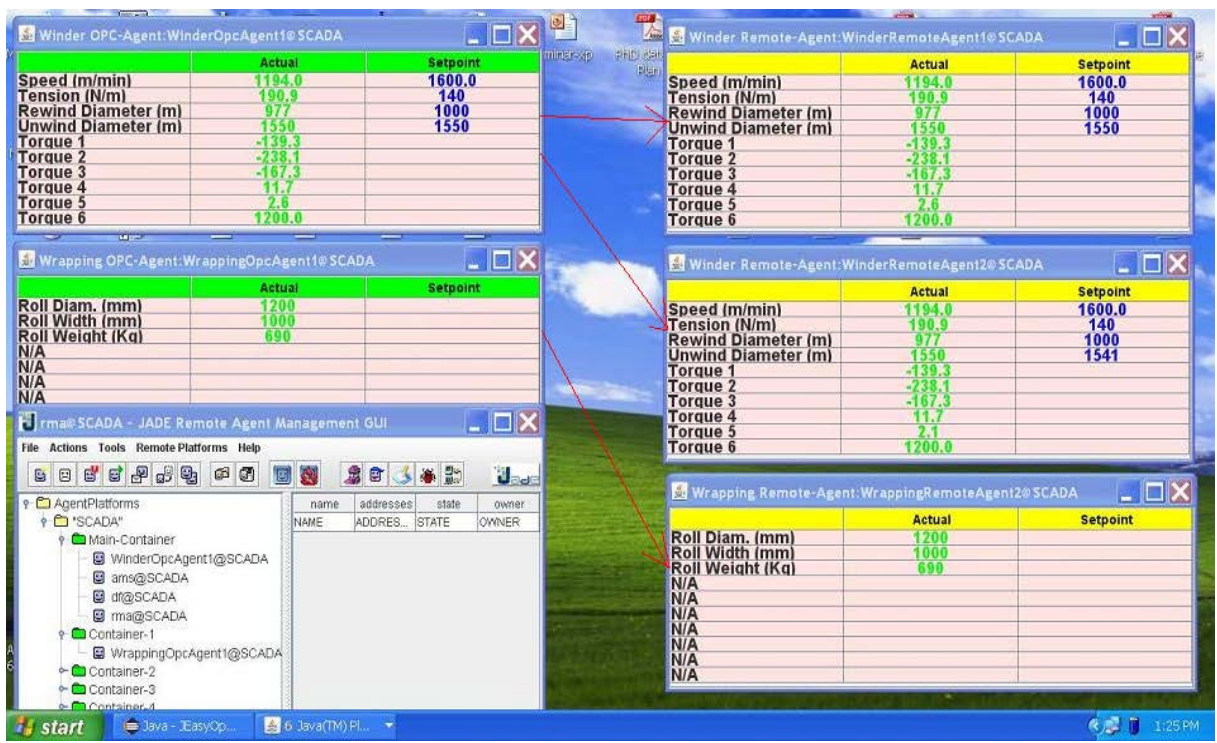

**Figure 14.** Running the concrete example: Two OPC-Agents (left side) and three normal Jade Agents (right side).

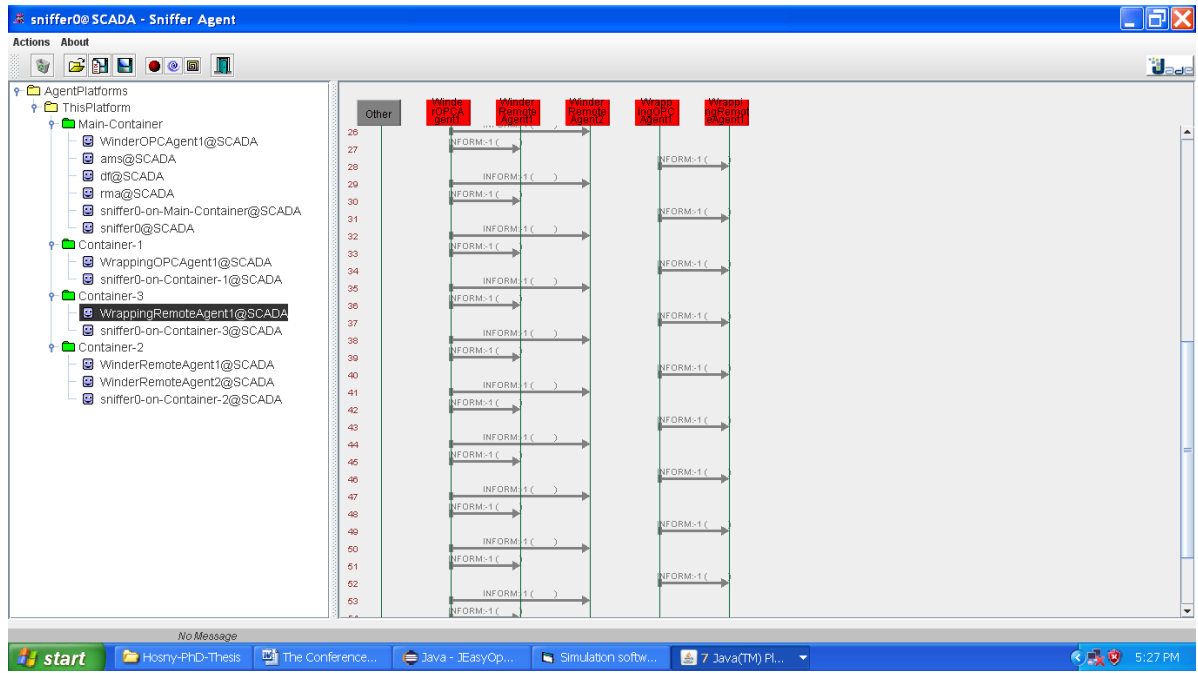

**Figure 15.** JADE's built-in sniffer tool, three normal jade agents, two of them subscribe to Winder OPCAgent and one subscribes to Wrapping OPCAgent.

## 6. Conclusion

Multi-agent systems propose solutions to highly distributed problems in dynamic and open computational domains. SCADA is one of these systems which are highly distributed, decentralized, open, and requires high degree of scalability. The agent-based approach should be adopted for developing flexible and scalable SCADA systems. In this research, the agent-based approach was adopted to develop a simple, flexible, and interoperable SCADA system. The achieved interoperability is realized by the adoption of the OPC technology for process communication, by this way it is possible to run the developed system with any type of control systems (*i.e.* PLC, DCS, CNC, etc.) independent of any vendor, and the same Java-OPC interface will be used with only some few modifications. By flexible, we mean that the system is able to adapt its dynamic working environment. The applicability of the agent-based approach is demonstrated by developing a practical project carried out in a paper mill. Furthermore, the proposed approach is cost-effective compared with custom SCADA packages. The future work will be the adoption of the agents' technology for building large-scale and highly distributed SCADA systems.

## Acknowledgements

Hosny Abbas thanks the managers of Quena Paper Company (Egypt), Mr. Shazely Abd-El Azeem (General Manager) and Mr. Abd-El Hameid Omar (Manager of Automation Sector) for their encouragement and their acceptance to allow us to carry out this research in the company at their responsibilities and supervisory.

## Appendix A: Java-OPC Interaction

The JEasyOPC project is established on LGPL (GNU Library or Lesser General Public License). The project can be downloaded from SourceForge.net as actual release (2.xx.xx) or night revision from SVN repository. The project is built on Eclipse 3.2.x Open Source IDE. The release is distributed as a zip-file for a quick download. In a zip-file, there are these important directories and files:

- *jeasyopc.jar*: the final library for usage in your application.
- *src.jar*: the source of library for a preview of library classes.
- *eclipse-project\JEasyOpc.zip*: zip-file with whole JEasyOPC project for Eclipse. There are all examples, JUnit tests, all sources!
- *Doc*: the directory includes documentation.
- *Resources*: the configuration files of JEasyOPC library. These resources have to be included in CLASSPATH of your project. There are all important information about usage of logging, internationalization and dll-library path (property library.path).

**Interface with OPC servers:**
Import javafish.clients.opc.JCustomOpc;
Import javafish.clients.opc.JEasyOpc;
Import javafish.clients.opc.JOpc;
Import javafish.clients.opc.asynch.AsynchEvent;
Import javafish.clients.opc.asynch.OpcAsynchGroupListener;
Import javafish.clients.opc.browser.JOpcBrowser;
Import javafish.clients.opc.component.OpcGroup;
Import javafish.clients.opc.component.OpcItem;
*//Initialization*
jopc_meas= new JEasyOpc(hostName, serverName, groupName+ "_meas");
*//Reading*
gotItem= jopc_meas.synchReadItem(statusGroup, (OpcItem)ItemToRead);
*//Writing*
jopc_command.synchWriteItem(commandGroup, item);

As an example, consider a JADE application contains two agents: one is connected through LAN to the OPC server and continuously read changed process variables from the server (call is OPC-Agent); the other is a remote agent interacts with the OPC-Agent by message passing to get the latest changed process variables. The following Java program presents a possible implementation of the OPC-Agent (to save paper size, exceptions handling is not included in the code):

```
package javafish.clients.opc;
// Jade Imports
import jade.core.*;
import jade.core.behaviours.TickerBehaviour;
import jade.lang.acl.ACLMessage;
// JEasyOPC imports
import javafish.clients.opc.component.*;
import javafish.clients.opc.exception.*;
public class Simple_OPC_Agent extends Agent
{
// OPC Declarations
private JEasyOpc jopc;
private OpcGroup group;
private OpcItem item1,item2;
private OpcItem responseItem1,responseItem2;
// Agent Setup function
protected void setup() {
// connecting to Siemens OPC.SimaticNet OPC server
jopc = new JEasyOpc("localhost", "OPC.SimaticNET", "JOPC1");
```

```
JOpc.coInitialize();
// OPC Group Creation
item1 = new OpcItem("s7:[@LOCALSERVER]db1,w0", true, "");
item2 = new OpcItem("s7:[@LOCALSERVER]db1,w2", true, "");
group = new OpcGroup("group1", true, 400, 0.0f);
group.addItem(item1);
group.addItem(item2);
jopc.addGroup(group);
// Starting the OPC Server
jopc.start();
addBehaviour(new TickerBehaviour(this, 1500) {
protected void onTick() {
responseItem1 = jopc.synchReadItem(group, item1);
responseItem2 = jopc.synchReadItem(group, item2);
System.out.println("item1="+ responseItem1.getValue());
System.out.println("item2="+ responseItem2.getValue());
ACLMessage msg1=new ACLMessage(ACLMessage.INFORM);
// Process the message
msg1.addReceiver(new AID("H2", AID.ISLOCALNAME));
msg1.setLanguage("English");
msg1.setContent(responseItem1.getValue()+responseItem2.getValue());
send(msg1);
}});}
```

In this example the OPC server is supposed to be *OPC. SimaticNET*, which is used to connect to Siemens control systems. Also the OPC server is supposed to be installed on the same machine as the sample client application which means that the communication between the client and the server takes place through COM, if the client and server are hosted by different machines then DCOM will be used. The remote agent is a simple Jade agent required to interact with the OPC-Agent by massage passing. First it might send only one starting message to the OPC-Agent and continuously, through a cyclic behavior, it receives messages from the OPC-Agent and retrieves the OPC data from these messages and prints them to the console. For more information about the application of JEasyOPC, interested readers are invited to read the documents and manuals of JEasyOPC.